\documentclass[9pt]{book}

\usepackage[dvips]{graphicx,color}
\usepackage{makeidx,universe}

\makeauthorindex
\BookTitle{The proceedings of the Physics of Accreting Compact Binaries}

\CopyRight{\copyright 2010 by Universal Academy Press, Inc.}

\begin{document} %%%%%%%%%%%%%%%

%\tableofcontents
\pagenumbering{arabic}

\chapter{%
%%%%%%%%%% <===== TITLE of the contribution
%%%%%%%%%%% The first letter of each word should be capital letter.
Spectral Analyses Of The Hottest White Dwarfs:\\
Grids Of Spectral Energy Distributions\\ 
For Extremely Hot, Compact Stars\\
In The Framework Of The Virtual Observatory
}

\author{\raggedright \baselineskip=10pt%
{\bf Thomas Rauch,
Ellen Ringat,
and 
Klaus Werner}\\ %%%%%% <== Authors
{\small \it %
Institute for Astronomy and Astrophysics,
Kepler Center for Astro and Particle Physics,
Eberhard Karls University,
Sand 1,
D-72076 T\"ubingen,
Germany}
}

%**************************
% Please note:
% One \AuthorContents{} is necessary
%    for EACH CONTRIBUTION, for the contents page and
% One \AuthorIndex{} is necessary
%    for EACH AUTHOR, for the index.
%**************************

\AuthorContents{Thomas Rauch, Ellen Ringat, Klaus Werner} %%%%%%% <=== It is the data for CONTENTS. Please enter all author's name that should be initialized.

\AuthorIndex{Rauch}{T.} %%%%%%% <=== It is the data for AUTHOR INDEX. Please enter a author's name that should be initialized.

\AuthorIndex{Ringat}{E.} %%%%%%% <=== It is the data for AUTHOR INDEX. Please enter a author's name that should be initialized.

\AuthorIndex{Werner}{K.} %%%%%%% <=== It is the data for AUTHOR INDEX. Please enter a author's name that should be initialized.

     \baselineskip=10pt
     \parindent=10pt

\section*{Abstract} %%%%%%%%%%%%%%%

Present X-ray missions like \emph{Chandra} and XMM-\emph{Newton} provide
high-resolution and high-S/N observations of extremely hot 
white dwarfs, e.g\@. burst spectra of novae. Their analysis
requires adequate Non-LTE model atmospheres. 

  The T\"ubingen Non-LTE Model-Atmosphere Package \emph{TMAP}
can calculate such model atmospheres and spectral energy
distributions at a high level of sophistication. 

  In the framework of the Virtual Observatory, the
German Astrophysical Virtual Observatory (\emph{GAVO}) offers
\emph{TheoSSA}, a \emph{Virtual Observatory} (\emph{VO}) service 
that provides easy access to theoretical SEDs. 

  We present a new grid of SEDs, that is calculated in
the parameter range of novae and supersoft X-ray sources.

\section{Introduction} %%%%%%%%%%%%%%%
\label{sect:introduction}

In the last two decades, the  
T\"ubingen Non-LTE Model-Atmosphere Package 
(\emph{TMAP}\footnote{http://astro.uni-tuebingen.de/\raisebox{.2em}{{\tiny $\sim$}}rauch/TMAP/TMAP.html}, \cite{werneretal2003,rd2003})
was successfully employed to perform spectral analyses of
hot, compact stars based on observations from the infrared
to the X-ray wavelength range 
(see, e.g\@. \cite{jahnetal2007, rauchetal2007}).

Many identifications of until then unidentified lines 
and subsequent determinations of the respective element abundances 
in recent years, that were based on
high-resolution and high-S/N observations obtained with
\emph{FUSE}\footnote{Far Ultraviolet Spectroscopic Explorer}
and HST/\emph{STIS}\footnote{Space Telescope Imaging Spectrograph}
(e.g\@. 
Ne\,{\sc vii}  \cite{werneretal2004b},
F\,{\sc vi}    \cite{wrk2005}, 
Ar\,{\sc vii}  \cite{wrk2007a}, 
Ne\,{\sc viii} \cite{wrk2007b}, 
Ca\,{\sc x}    \cite{wrk2008}, and
Fe\,{\sc x}    \cite{wrk2010}),
are the direct consequence of
a continuous improvement of the \emph{TMAP} code 
and its input atomic data.

\emph{TMAP} Non-LTE models also provide SEDs
for hot, compact stars. One important application are theoretical
ionizing stellar fluxes, that are a necessary input for state-of-the-art
photoionization models of nebulae, like the 3-D code \emph{MOCASSIN} \cite{ercolanoetal2005}
and \emph{CLOUDY} \cite{ferlandetal1998}. Both are able to deal with
standard \emph{TMAP} flux tables. 
However, the way into the 21$^\mathrm{st}$ century, i.e\@. towards the use of
reliable theoretical SEDs, e.g\@. in the planetary-nebulae community, was
long -- and still there is a temptation to use the ``so-easy-to-calculate''
blackbodies, that are of course only a very coarse approximation of any star. 
The blackbody flux maximum is in general at lower energy compared to a
stellar spectrum of the same effective temperature ($T_\mathrm{eff}$) and
their peak intensity is lower \cite{rauch1997,rauch2003}. We describe how easy it is
today to retrieve Non-LTE SEDs via the \emph{VO} service
\emph{TheoSSA} in Sect.~\ref{sect:theossa}

\emph{TMAP} can, in general, calculate models up to $T_\mathrm{eff}$ of about 10\,MK.
In applications to novae \cite{rauchetal2010}, 
\emph{TMAP} still lacks some physics, most notably the velocity fields. 
However, it is a flexible and robust tool for 
the determination of basic parameters like $T_\mathrm{eff}$,
for line identifications \cite{rauchetal2008}, 
and to derive a reliable range of abundances in the white-dwarf atmosphere.
We briefly describe \emph{TMAP} and the atomic data that
was considered for the model-atmosphere calculation in Sect.~\ref{sect:tmap}

\section{\emph{TMAP} Non-LTE Models And Atomic Data}
\label{sect:tmap}

\emph{TMAP} is capable to calculate plane-parallel and spherical,
chemically homogeneous
Non-LTE model atmospheres 
in radiative and hydrostatic equilibrium. It
considers opacities of all species from hydrogen to nickel.
Many studies (Sect.~\ref{sect:introduction}) have shown that
\emph{TMAP} is a proven tool in spectral analysis of optical,
UV, and X-ray \cite{werneretal2004a}. 

The main limitation that we encounter now is the lack of 
reliable atomic and line-broadening data.
Going to higher resolution and S/N in the observations reveals 
uncertainties in atomic data even for the most abundant species. 
There are always unidentified lines in UV observations 
\cite{jahnetal2007}, that most likely stem from highly ionized 
light metals like e.g\@. neon or magnesium.

In the framework of \emph{GAVO}, we have set up the
T\"ubingen Model-Atom Database 
(\emph{TMAD}\footnote{http://astro.uni-tuebingen.de/\raisebox{.2em}{{\tiny $\sim$}}TMAD/TMAD.html}).
In contains ready-to-use model atoms in the \emph{TMAP} format,
that comprise most recent atomic data. 
For illustration, Fig.~1 shows the complexity of our neon model atom.
The \emph{TMAD} model atoms may
be used by any other model-atmosphere code -- provided that
a suitable interface exists.

\begin{figure}[ht!]
\resizebox{0.91\hsize}{!}{\includegraphics{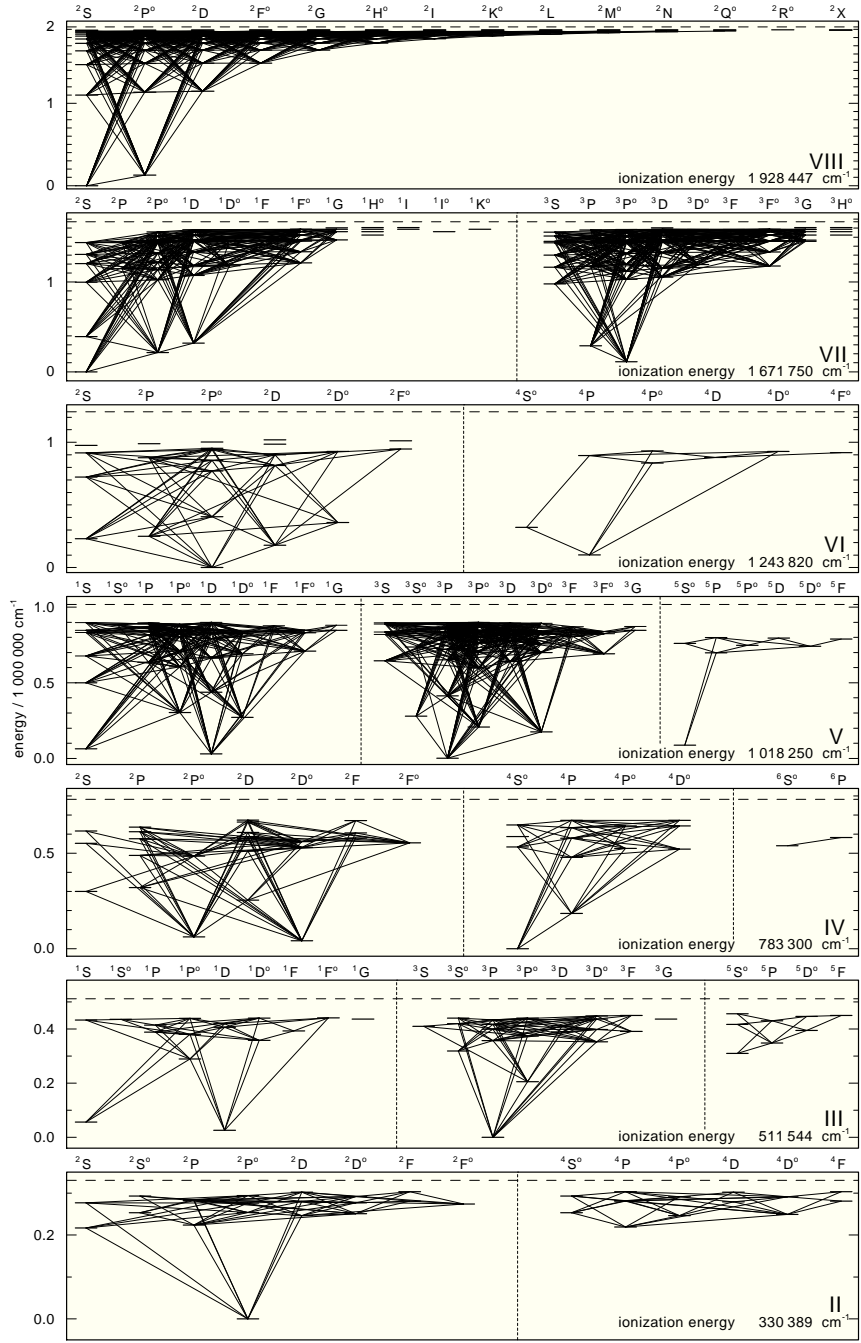}}\centering
\caption{Grotrian diagrams of the Ne {\sc iii} - {\sc viii} model ions 
         that are provided by \emph{TMAD}.}
\label{fig:grotrian}
\end{figure}

Within our recent analysis of the compact component in nova V4743\,Sgr \cite{rauchetal2010}, 
we calculated extended grids of model atmospheres 
with $T_\mathrm{eff} = 0.45 - 1.05\,\mathrm{MK}$ and $\Delta T_\mathrm{eff} = 0.01\,\mathrm{MK}$
and $\log g = 9$. The presently available abundances are summarized in Tab.~\ref{tab:abundances}
See \cite{rauchetal2010} as an example of the application of these
grids for spectral analyses.
Similar grids with lower $\log g$ are currently calculated.
All these  grids are available now via \emph{TheoSSA} (Sect.~3),
and, already converted to 
atables\footnote{http://astro.uni-tuebingen.de/$^\sim$rauch/TMAF/flux\_HHeCNONeMgSiS\_gen.html} 
for the use with 
\emph{XSPEC}\footnote{http://heasarc.gsfc.nasa.gov/docs/xanadu/xspec}.
As an example, Fig.~2 shows SEDs from model grid 003.

\begin{table}[ht!]
\small
\caption{Elements and their abundances (given as log\,[abundance / solar abundance],
         solar values from \cite{asplundetal2009})
         that are considered in our model grids (003 $-$ 011).
         Fe is a generic model atom \cite{rd2003} that includes the elements Ca $-$ Ni.} 
\label{tab:abundances}
\begin{center}
\begin{tabular}{cr@{.}lr@{.}lr@{.}lr@{.}lr@{.}lr@{.}lr@{.}lr@{.}lr@{.}l} 
\hline
\hline
\noalign{\smallskip}
  &   
\multicolumn{2}{c}{003} &     
\multicolumn{2}{c}{004} &     
\multicolumn{2}{c}{005} &     
\multicolumn{2}{c}{006} &     
\multicolumn{2}{c}{007} &     
\multicolumn{2}{c}{008} &     
\multicolumn{2}{c}{009} &    
\multicolumn{2}{c}{010} &    
\multicolumn{2}{c}{011}  \\
\cline{2-19}
\noalign{\smallskip}
  H     & -0&688 & -0&683 & -0&677 & -0&673 & -0&672 & -0&671 & -0&670 & -0&670 & -0&669 \\
  He    &  0&382 &  0&387 &  0&393 &  0&397 &  0&398 &  0&399 &  0&400 &  0&401 &  0&401 \\
  C     & -1&513 & -1&073 & -0&772 & -0&675 & -0&596 & -0&529 & -0&471 & -0&420 & -0&374 \\
  N     &  1&803 &  1&678 &  1&460 &  1&159 &  1&062 &  0&937 &  0&761 &  0&460 &  0&159 \\
  O     &  1&528 &  1&533 &  1&538 &  1&543 &  1&544 &  1&544 &  1&545 &  1&546 &  1&547 \\
  Ne    & -0&474 & -0&469 & -0&464 & -0&459 & -0&459 & -0&458 & -0&457 & -0&456 & -0&456 \\
  Mg    & -0&454 & -0&450 & -0&444 & -0&439 & -0&439 & -0&438 & -0&437 & -0&436 & -0&436 \\
  Si    &  0&167 &  0&172 &  0&178 &  0&182 &  0&183 &  0&184 &  0&185 &  0&186 &  0&186 \\
  S     & -1&583 & -1&578 & -1&573 & -1&568 & -1&567 & -1&567 & -1&566 & -1&565 & -1&565 \\
  Fe    &  0&828 &  0&833 &  0&838 &  0&843 &  0&843 &  0&844 &  0&845 &  0&846 &  0&846 \\
\hline
\end{tabular}
\end{center}
\end{table}

\begin{figure}[ht!]
\includegraphics{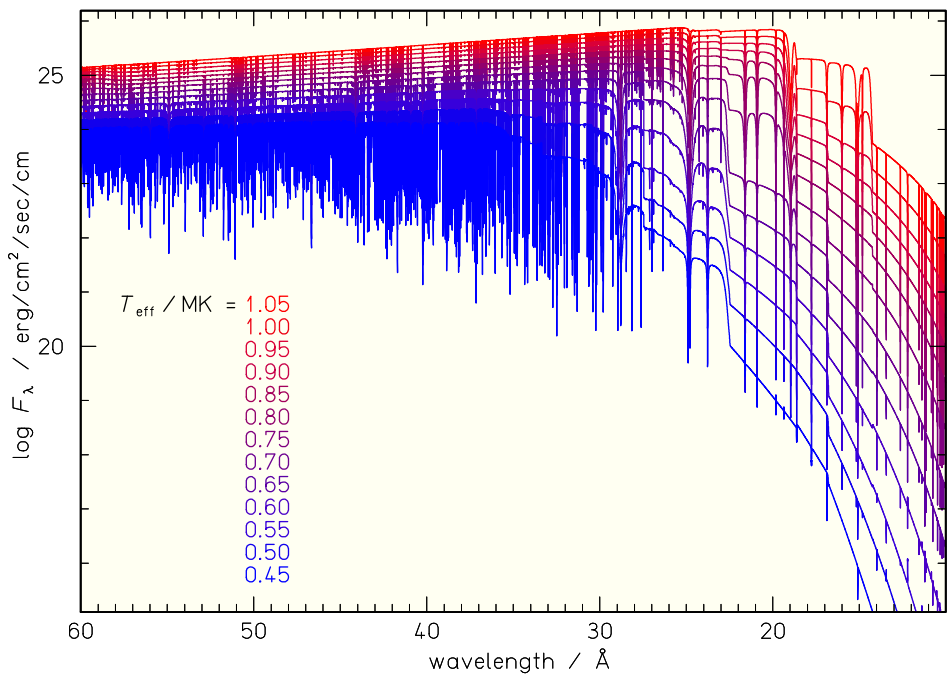}\centering
\caption{Selected ($\Delta T_\mathrm{eff} = 0.05\,\mathrm{MK}$) theoretical SEDs from model grid 003 (Tab.~1).}
\label{fig:sed}
\end{figure}

\setcounter{footnote}{1}

\section{\emph{TheoSSA} -- Theoretical Stellar Spectra On Demand}
\label{sect:theossa}

In the framework of the \emph{VO}\footnote{http://www.ivoa.net}.
\emph{TheoSSA}\footnote{Theoretical Simple Spectra Access} is a
registered service, 
provided by \emph{GAVO}\footnote{http://www.g-vo.org}.
SEDs are easily accessible in \emph{VO} compliant form via the 
\emph{TheoSSA} WWW interface\footnote{http://vo.ari.uni-heidelberg.de/ssatr-0.01/TrSpectra.jsp?}.
The SEDs are available at three levels:\\
   1) \emph{fast and easy}: pre-calculated SED grids that span generally over a wide range of $T_\mathrm{eff}$ 
   ($50 - 190\,\mathrm{kK}$) 
   and surface gravity 
   ($\log g = 5 - 9$)
   for different chemical compositions, 
   e.g., pure H, pure He, H+He, He+C+N+O, C+O+Ne+Mg,
   H -- Ca \cite{rauch1997}, and H -- Ni \cite{rauch2003}.\\
   2) \emph{individual}: model atmospheres based on standard model atoms -- neither profound knowledge of theory nor
   experience with the software is here a pre-requisite. The photospheric parameters
   $T_{\mathrm{eff}}$, $\log g$, and mass fractions $\{X_{\mathrm{i}}\}$
   for $i \in \left[\mathrm{H, He, C, N, O}\right]$ can be adjusted in order to improve the fit to the observation.
   This is performed via \emph{TMAW}, a WWW service within \emph{TheoSSA}.\\
   3) \emph{experienced}: observers and theoreticians, who want to compare e.g\@. their own simulations with
   results of \emph{TMAP}, the creation and upload of own atomic-data files is possible.
   \emph{TMAD} provides model atoms which are suited for the use by {\sc TMAP}. These may be adjusted for an
   individual object.

\begin{figure*}[ht!]
{\tiny
\begin{picture}(17.0,17.5)
\thicklines
%
%
% left panel
%
\put( 4.5,17.0){\oval( 1.0, 1.0)[tl]}
\put(10.5,17.0){\oval( 1.0, 1.0)[tr]}
\put( 4.5,15.5){\oval( 1.0, 1.0)[bl]}
\put(10.5,15.5){\oval( 1.0, 1.0)[br]}
\put( 4.5,17.5){\line( 1, 0){ 6.0}}
\put( 4.5,15.0){\line( 1, 0){ 6.0}}
\put( 4.0,15.5){\line( 0, 1){ 1.5}}
\put(11.0,15.5){\line( 0, 1){ 1.5}}
\put( 4.0,16.5){\makebox( 7.0, 1.0)[c]{
{{\color{green}\sc TMAW} request via WWW interface:}
}}
\put( 4.0,16.0){\makebox( 7.0, 1.0)[c]{
{\color{red}$T_{\mathrm{eff}}$, $\log g$, $\{X_{\mathrm{i}}\}$}
}}
\put( 4.0,15.5){\makebox( 7.0, 1.0)[c]{
{standard flux table}
}}
\put( 4.0,15.0){\makebox( 7.0, 1.0)[c]{
{individual flux table: $\lambda$ interval, resolution}
}}
\put( 7.5,15.0){\vector( 0,-1){ 1.0}}
\put( 5.0,13.5){\oval( 1.0, 1.0)[tl]}
\put(10.0,13.5){\oval( 1.0, 1.0)[tr]}
\put( 5.0,12.5){\oval( 1.0, 1.0)[bl]}
\put(10.0,12.5){\oval( 1.0, 1.0)[br]}
\put( 5.0,14.0){\line( 1, 0){ 5.0}}
\put( 5.0,12.0){\line( 1, 0){ 5.0}}
\put( 4.5,12.5){\line( 0, 1){ 1.0}}
\put(10.5,12.5){\line( 0, 1){ 1.0}}
\put( 4.0,12.0){\makebox( 7.0, 3.0)[c]{
check of \emph{GAVO} database:
}}
\put( 4.0,11.5){\makebox( 7.0, 3.0)[c]{
are requested parameters matched
}}
\put( 4.0,11.0){\makebox( 7.0, 3.0)[c]{
within tolerance limits?
}}
\put(10.5,13.0){\vector( 1, 0){ 3.0}}
\put(13.5,13.0){\vector(-1, 0){ 3.0}}
\put( 7.50,12.00){\line( 0,-1){ 0.5}}
\put( 4.75,11.50){\line( 1, 0){ 5.5}}
\put( 4.75,11.50){\vector( 0,-1){ 0.5}}
\put(10.25,11.50){\vector( 0,-1){ 0.5}}
%
%
% left
%
\put( 3.0,10.5){\oval( 1.0, 1.0)[tl]}
\put( 6.5,10.5){\oval( 1.0, 1.0)[tr]}
\put( 3.0,10.0){\oval( 1.0, 1.0)[bl]}
\put( 6.5,10.0){\oval( 1.0, 1.0)[br]}
\put( 3.0,11.0){\line( 1, 0){ 3.5}}
\put( 3.0, 9.5){\line( 1, 0){ 3.5}}
\put( 2.5,10.0){\line( 0, 1){ 0.5}}
\put( 7.0,10.0){\line( 0, 1){ 0.5}}
\put( 2.0,10.0){\makebox( 5.5, 1.0)[c]{
{\color{blue}yes}:
}}
\put( 2.0, 9.5){\makebox( 5.5, 1.0)[c]{
offer existing model
}}
%
%
% right
%
\put( 8.5,10.5){\oval( 1.0, 1.0)[tl]}
\put(12.0,10.5){\oval( 1.0, 1.0)[tr]}
\put( 8.5,10.0){\oval( 1.0, 1.0)[bl]}
\put(12.0,10.0){\oval( 1.0, 1.0)[br]}
\put( 8.5,11.0){\line( 1, 0){ 3.5}}
\put( 8.5, 9.5){\line( 1, 0){ 3.5}}
\put( 8.0,10.0){\line( 0, 1){ 0.5}}
\put(12.5,10.0){\line( 0, 1){ 0.5}}
\put( 7.5,10.0){\makebox( 5.5, 1.0)[c]{
{\color{red}no}:
}}
\put( 7.5, 9.5){\makebox( 5.5, 1.0)[c]{
calculate new model
}}
\put(12.5,10.25){\vector( 1, 0){ 1.0}}
\put( 4.75, 9.50){\line( 0,-1){ 0.5}}
\put( 2.75, 9.00){\line( 1, 0){ 4.0}}
\put( 2.75, 9.00){\vector( 0,-1){ 0.5}}
\put( 6.75, 9.00){\vector( 0,-1){ 0.5}}
%
%
% left
%
\put( 1.5, 8.0){\oval( 1.0, 1.0)[tl]}
\put( 4.0, 8.0){\oval( 1.0, 1.0)[tr]}
\put( 1.5, 6.0){\oval( 1.0, 1.0)[bl]}
\put( 4.0, 6.0){\oval( 1.0, 1.0)[br]}
\put( 1.5, 8.5){\line( 1, 0){ 2.5}}
\put( 1.5, 5.5){\line( 1, 0){ 2.5}}
\put( 1.0, 6.0){\line( 0, 1){ 2.0}}
\put( 4.5, 6.0){\line( 0, 1){ 2.0}}
\put( 0.5, 7.5){\makebox( 4.5, 1.0)[c]{
{\color{blue}accept}
}}
\put( 0.5, 6.5){\makebox( 4.5, 1.0)[c]{
{retrieve flux tables,}
}}
\put( 0.5, 6.0){\makebox( 4.5, 1.0)[c]{
{on-the-fly products}
}}
\put( 0.5, 5.5){\makebox( 4.5, 1.0)[c]{
{from \emph{GAVO} database}
}}
\put(13.50, 6.50){\vector(-1,0){ 9.0}}
%
%
% right
%
\put( 5.5, 8.0){\oval( 1.0, 1.0)[tl]}
\put( 8.0, 8.0){\oval( 1.0, 1.0)[tr]}
\put( 5.5, 7.5){\oval( 1.0, 1.0)[bl]}
\put( 8.0, 7.5){\oval( 1.0, 1.0)[br]}
\put( 5.5, 8.5){\line( 1, 0){ 2.5}}
\put( 5.5, 7.0){\line( 1, 0){ 2.5}}
\put( 5.0, 7.5){\line( 0, 1){ 0.5}}
\put( 8.5, 7.5){\line( 0, 1){ 0.5}}
\put( 4.5, 7.5){\makebox( 4.5, 1.0)[c]{
{\color{red}request}
}}
\put( 4.5, 7.0){\makebox( 4.5, 1.0)[c]{
{\color{red} exact $T_{\mathrm{eff}}$, $\log g$, $\{X_{\mathrm{i}}\}$}
}}
\put( 8.50, 7.75){\line( 1, 0){ 1.75}}
\put(10.25, 7.75){\vector( 0, 1){ 1.75}}
%
%
% right panel
%
\put(14.0,13.5){\oval( 1.0, 1.0)[tl]}
\put(16.0,13.5){\oval( 1.0, 1.0)[tr]}
\put(14.0, 5.5){\oval( 1.0, 1.0)[bl]}
\put(16.0, 5.5){\oval( 1.0, 1.0)[br]}
\put(14.0,14.0){\line( 1, 0){ 2.0}}
\put(14.0, 5.0){\line( 1, 0){ 2.0}}
\put(13.5, 5.5){\line( 0, 1){ 8.0}}
\put(16.5, 5.5){\line( 0, 1){ 8.0}}
\put(13.5,12.0){\makebox( 3.0, 3.0)[c]{
\emph{GAVO} database
}}
\put(13.5,11.5){\makebox( 3.0, 3.0)[c]{
{\color{red}$T_{\mathrm{eff}}$, $\log g$, $\{X_{\mathrm{i}}\}$}
}}
\put(14.2,11.75){\oval( 1.0, 1.0)[tl]}
\put(15.8,11.75){\oval( 1.0, 1.0)[tr]}
\put(14.2,11.00){\oval( 1.0, 1.0)[bl]}
\put(15.8,11.00){\oval( 1.0, 1.0)[br]}
\put(14.2,12.25){\line( 1, 0){ 1.6}}
\put(14.2,10.50){\line( 1, 0){ 1.6}}
\put(13.7,11.00){\line( 0, 1){ 0.75}}
\put(16.3,11.00){\line( 0, 1){ 0.75}}
\put(13.5,10.5){\makebox( 3.0, 3.0)[c]{
{\color{green}ARI}
}}
\put(13.5,10.0){\makebox( 3.0, 3.0)[c]{
{\color{blue}meta data}
}}
\put(13.5, 9.5){\makebox( 3.0, 3.0)[c]{
{VO services}
}}
\put(15.0,10.50){\vector( 0,-1){ 0.75}}
\put(14.2, 9.25){\oval( 1.0, 1.0)[tl]}
\put(15.8, 9.25){\oval( 1.0, 1.0)[tr]}
\put(14.2, 6.00){\oval( 1.0, 1.0)[bl]}
\put(15.8, 6.00){\oval( 1.0, 1.0)[br]}
\put(14.2, 9.75){\line( 1, 0){ 1.6}}
\put(14.2, 5.50){\line( 1, 0){ 1.6}}
\put(13.7, 6.00){\line( 0, 1){ 3.25}}
\put(16.3, 6.00){\line( 0, 1){ 3.25}}
\put(13.5, 8.0){\makebox( 3.0, 3.0)[c]{
{\color{green}IAAT}
}}
\put(13.5, 7.5){\makebox( 3.0, 3.0)[c]{
{\color{blue}models}
}}
\put(13.5, 7.0){\makebox( 3.0, 3.0)[c]{
{atomic data}
}}
\put(13.5, 6.5){\makebox( 3.0, 3.0)[c]{
{frequency grids}
}}
\put(13.5, 6.0){\makebox( 3.0, 3.0)[c]{
{\color{blue}flux tables}
}}
\put(13.5, 5.5){\makebox( 3.0, 3.0)[c]{
\hbox{}\hspace{3.5mm}5 $-$ \hbox{}\hspace{1.2mm}2000\AA}}
\put(13.5, 5.0){\makebox( 3.0, 3.0)[c]{
\hbox{}\hspace{0.0mm}2000 $-$ \hbox{}\hspace{1.0mm}3000\AA}}
\put(13.5, 4.5){\makebox( 3.0, 3.0)[c]{
\hbox{}\hspace{0.2mm}3000 $-$ 55000\AA
}}
\end{picture}\vspace{-35mm}
\caption{Flow diagram of \emph{TheoSSA}.}
\label{fig:tmaw}
}
\end{figure*}
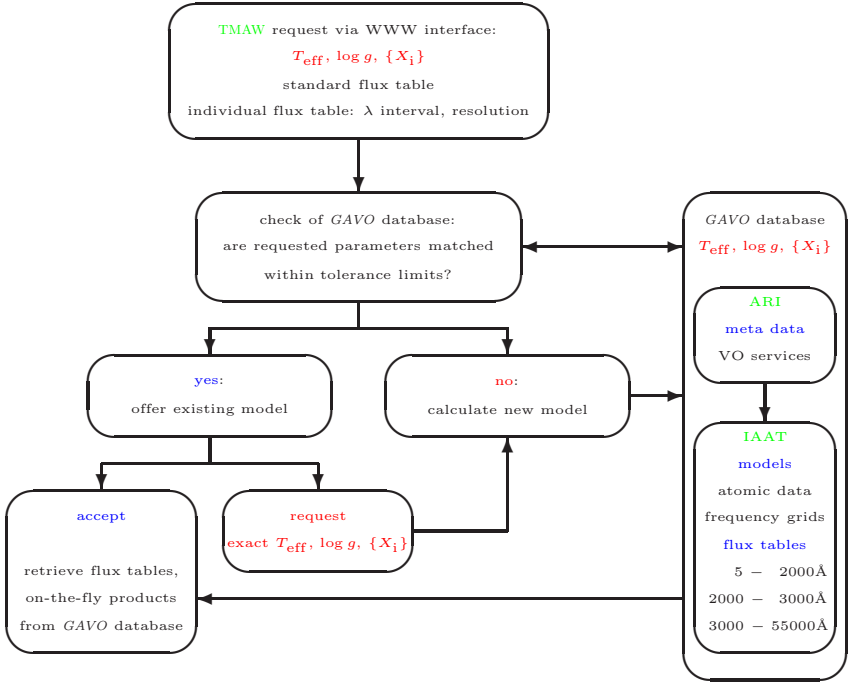

The usage of \emph{TheoSSA} is simple (Fig.~3). 
A \emph{VO} user submits a SED request to the \emph{GAVO} database. If a suitable model is available within tolerance
limits, this is offered. In case that the parameters are not exactly matched, the user
may decide to request a model with the exact parameters. \emph{TMAW} will start a model-atmosphere calculation at the
IAAT then. As soon as the model is converged, the SED is automatically ingested in the \emph{GAVO} database,
that is growing in time in this way,
and the user can retrieve the SED and various on-the-fly products from \emph{TheoSSA}.

The calculation of a single H+He+C+N+O model needs about one day on a presently available 64\,bit standard PC.
For the calculation of extended model grids in reasonable time, 
\emph{TheoSSA} makes use of compute resources provided by \emph{AstroGrid-D}\footnote{http://www.gac-grid.de/}.

\section{Conclusions}
\label{sect:conclusions}

Theoretical spectral energy distributions for extremely hot, compact stars
are available and to retrieve them via 
the \emph{GAVO} service \emph{TheoSSA} (Sect.~\ref{sect:theossa})
is easy -- use them! 

Spectral analysis by means of Non-LTE model-atmosphere techniques has for a long 
time been regarded as a domain of specialists.
With \emph{TheoSSA}, the access to individually calculated SEDs is as simple 
as the use of pre-calculated SEDs -- without detailed knowledge of the programme,
that is calculating in the background. 
However, the user has to be aware of the impact of metal opacities on the flux in the high-energy range 
\cite{rauch1997,rauch2003,rauchetal2010} and has to use
appropriate SEDs  for individual objects.
In case of doubt or for any question, please do not hesitate to contact
astro-tmaw@listserv.uni-tuebingen.de directly.

\clearpage

\paragraph{Acknowledgement}
TR is supported by the German Aerospace Center (DLR) under grant 05\,OR\,0806, 
and by a travel grant of the German Academic Exchange Service (DAAD).
ER is supported by the German Research Foundation (DFG) under grant WE\,1312/41-1.
We thank the \emph{GAVO} and \emph{AstroGrid-D} teams for support.
\vfill

%%%%%%%%%%%%%%%%%%%%%%%%%%%%%%%%%%
%% thebibliography environment %%
%%%%%%%%%%%%%%%%%%%%%%%%%%%%%%%%%

%%%%%%%%%%

\end{document}